\documentclass[prl,twocolumn,showpacs,preprintnumbers,amsmath]{revtex4}
\usepackage{graphicx}	
\usepackage{amssymb}	
\usepackage{amsmath}	
\usepackage{mathptm}	
\usepackage{multirow}
\usepackage{array}
\usepackage{hyperref}
\usepackage{color}	

\newcommand{\be}{\begin{equation}}
\newcommand{\ee}{\end{equation}}
\newcommand{\bea}{\begin{eqnarray}}
\newcommand{\eea}{\end{eqnarray}}

\begin{document}



\title{Biological Principles in Self-Organization of Young Brain - Viewed from Kohonen Model}

\author{T.~Pallaver$^{a,b}$, 
H.~Kr\"{o}ger$^{a}$\footnote{Corresponding author, Email:
hkroger@phy.ulaval.ca},
M.~Parizeau$^{c}$} 

\affiliation{
$^{a}$ {\small\sl D\'{e}partement de Physique, Universit\'{e} Laval, 
Qu\'{e}bec, Qu\'{e}bec G1K 7P4, Canada} \\
$^{b}$ {\small\sl Ecole Sup\'erieure d'Optique, Campus Polytechnique, RD 128, 91127 Palaiseau cedex, France} \\
$^{c}$ {\small\sl D\'{e}partement de G\'enie Electrique et de G\'enie Informatique, 
Universit\'{e} Laval, Qu\'{e}bec, Qu\'{e}bec G1K 7P4, Canada}
}

\date{\today}


\begin{abstract}

Variants of the Kohonen model are proposed to study biological principles of self-organization in a model of young brain. We suggest a function to measure aquired knowledge and use it to auto-adapt the topology of neuronal connectivity, yielding substantial organizational improvement relative to the standard model. In the early phase of organization with most intense learning, we observe that neural connectivity is of Small World type, which is very efficient to organize neurons in response to stimuli. In analogy to human brain where pruning of neural connectivity (and neuron cell death) occurs in early life, this feature is present also in our model, which is found to stabilize neuronal response to stimuli. 

\end{abstract}


\maketitle


{\it Introduction}. 
In physics the reductionist school of thought has long time prevailed, which postulates that the way of understanding laws of nature is by looking at its most elementary building blocks. This point of view has been questioned from the viewpoint of complexity, which says that in nature there are emergent phenomena which cannot be predicted from the laws of elementary constituents - for example the shape of a snow flake cannot be obtained from the laws of hydrogen and oxygen atoms \cite{Laughlin05}. 
A prime example of complexity is the self-organization of human brain. 
Experimental information on the organization of brain comes from observation of power laws $1/f^{\alpha}$ (with $\alpha \approx 1$) in the frequency power spectrum of electrical signals measured by electroencephalograms (EEG) \cite{Pritchard92,Robinson01} and by magnetoencephalograms (MEG) \cite{Novikov97}. Such power laws reflect a fractal temporal behavior \cite{Teich97}. Power laws have been found also in size distributions of epilectic seizures and neural avalanches \cite{Beggs03}, hinting to a fractal behavior in seize. 
To explain such power laws the sand pile model of self-organized criticality (SOC) has been devised \cite{Bak87,Rios99} and used to model EEG spectra \cite{Arcangelis06}. On the other hand, power laws in frequency have been observed in local field potentials (LFP) measured in cat parietal association cortex, which seem inconsistent with the SOC model \cite{Bedard06}. From the information theory point of view, however, it is not clear which information is transmitted in EEG signals or neural avalanches. \\
\indent Here we consider self-organization of young brain, based on the Kohonen model \cite{KohonenStand,KohonenNew}. It is a biological property that the organization of visual cortex in mammals occurs during a relative short time window shortly after birth during which neural connectivity is established \cite{Linsker,Malsburg}. Most remarkably, this is accompanied by genetically controlled neuron cell death and pruning of synaptic connections \cite{Pruning, Pruning2}. In brains of mature adults, functional magnetic resonance imaging (FMRI) has shown evidence for functional networks of correlated brain activity \cite{Eguiluz05}, which display {\it small world} and {\it scale free} architecture. {\it Small World Network} (SWN) neural connectivity has been found in a network of Hodgkin-Huxley neurons to give a fast and coherent system response \cite{Lago00}, in an associative memory model to improve restoration of perturbed memory patterns 
\cite{Bohland01} and to reduce learning time and error in a multi-layer 
learning model \cite{Simard05}. \\ 
\indent In this paper, three main issues are addressed:  
(i) we establish a link between network topology and flow of information, and show that biologically inspired auto-adaption of the Kohonen network improves organization;
(ii) we observe that SWN topology is present during most of the organizational phase;
(iii) we show that biologically observed pruning of synaptic connections in the early evolution of the network is beneficial for organization, while reconnections give an adverse effect.

\begin{figure}
\centering
\begin{tabular}[t]{l l}
\includegraphics[width=0.45\linewidth{}]{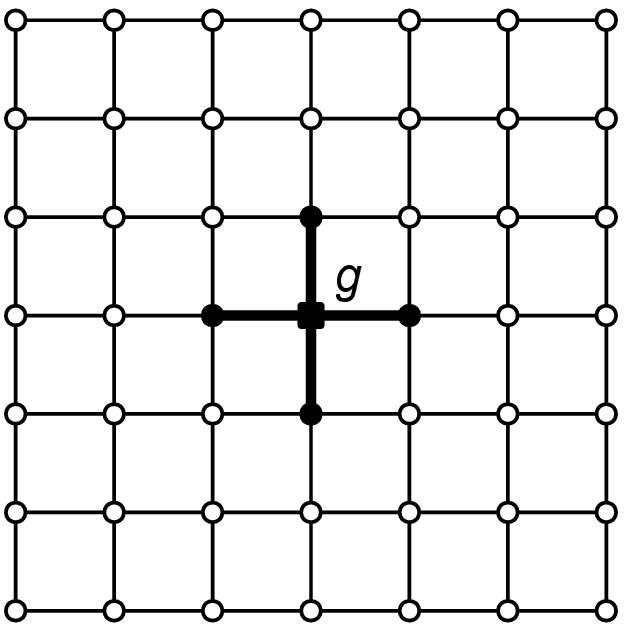} & 
\includegraphics[width=0.45\linewidth]{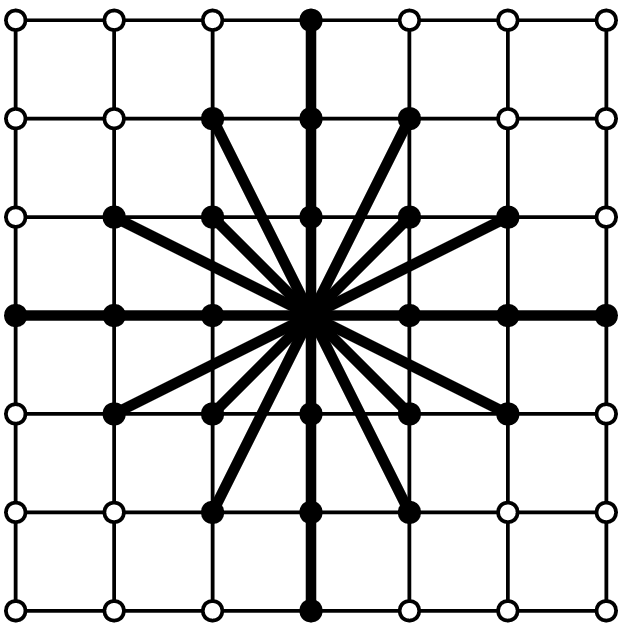} 
\end{tabular}
\caption{Initial regular grid map of neuron topology (neighbors): $V_{g}=1$ (left) and $V_{g}=3$ (right).}
\label{fig:grille10x10}
\end{figure}
\begin{figure}
\centering
\begin{tabular}[t]{c c}
\includegraphics[width=0.45\linewidth{}]{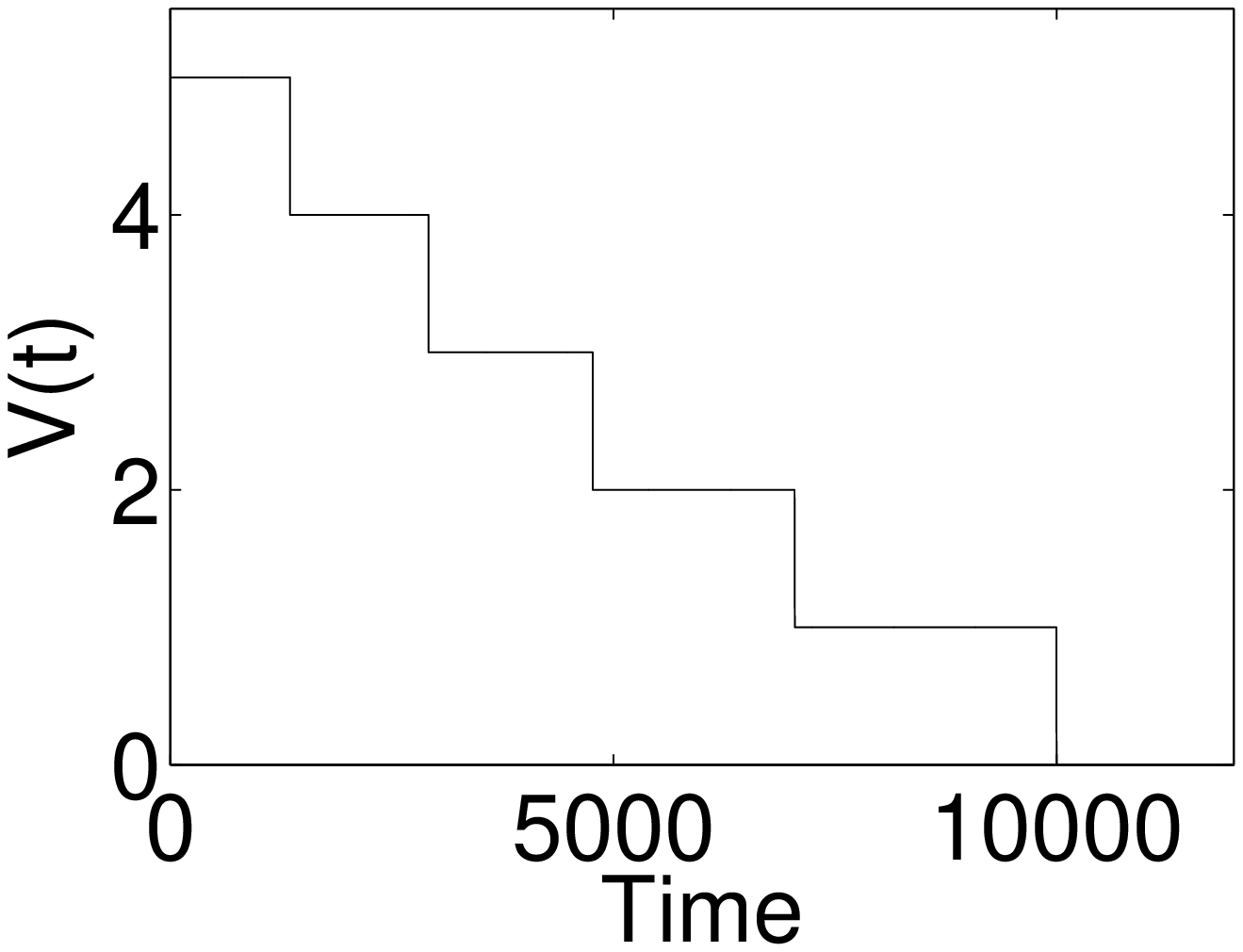} & 
\includegraphics[width=0.45\linewidth]{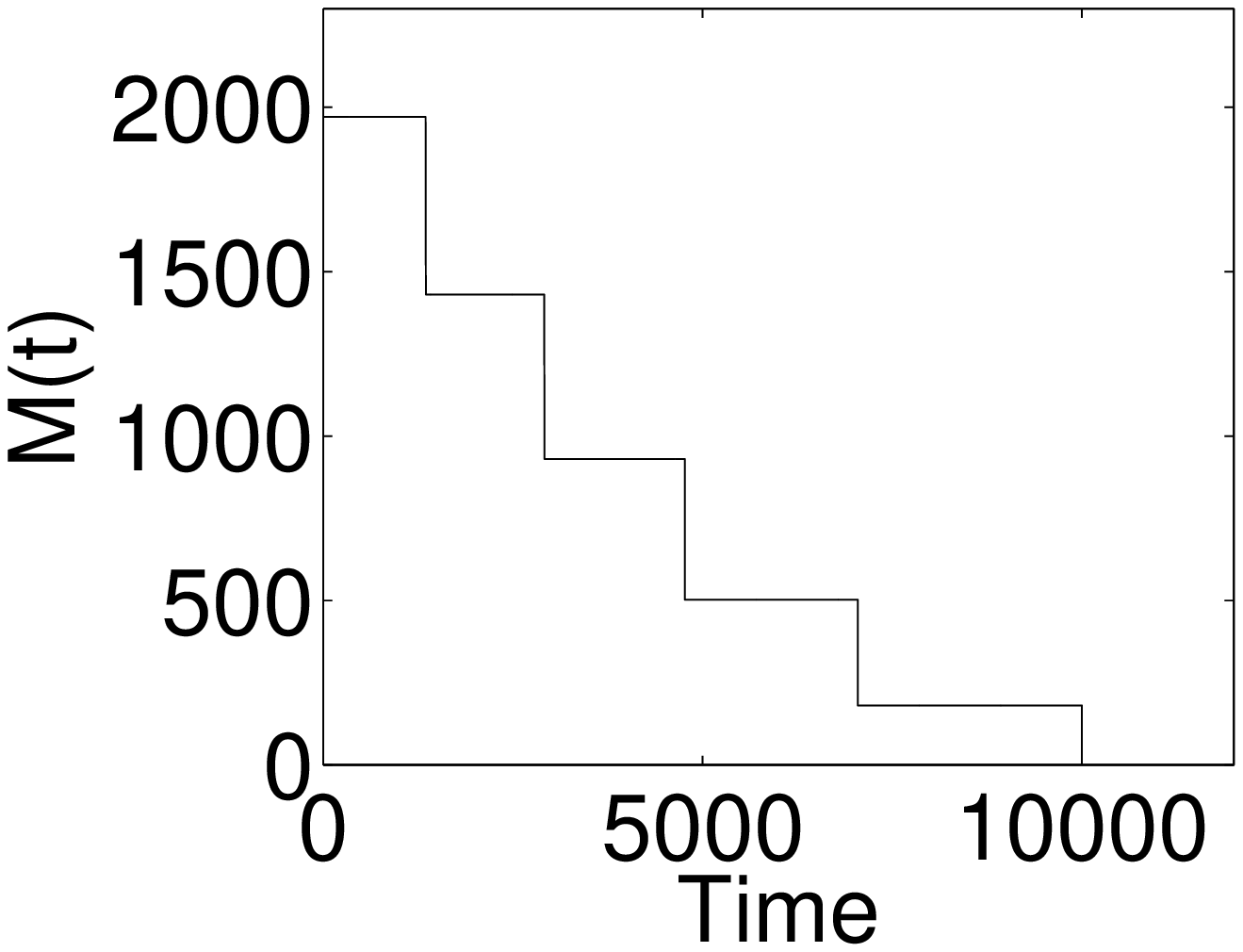}
\end{tabular}
\caption{Standard Kohonen. Relation between neighborhood order $V(t)$ and total number of connexions $M(t)$.}
\label{fig:correspondanceVK}
\end{figure}
\begin{figure}
\centering
\begin{tabular}[t]{c c}
\includegraphics[width=0.50\linewidth]{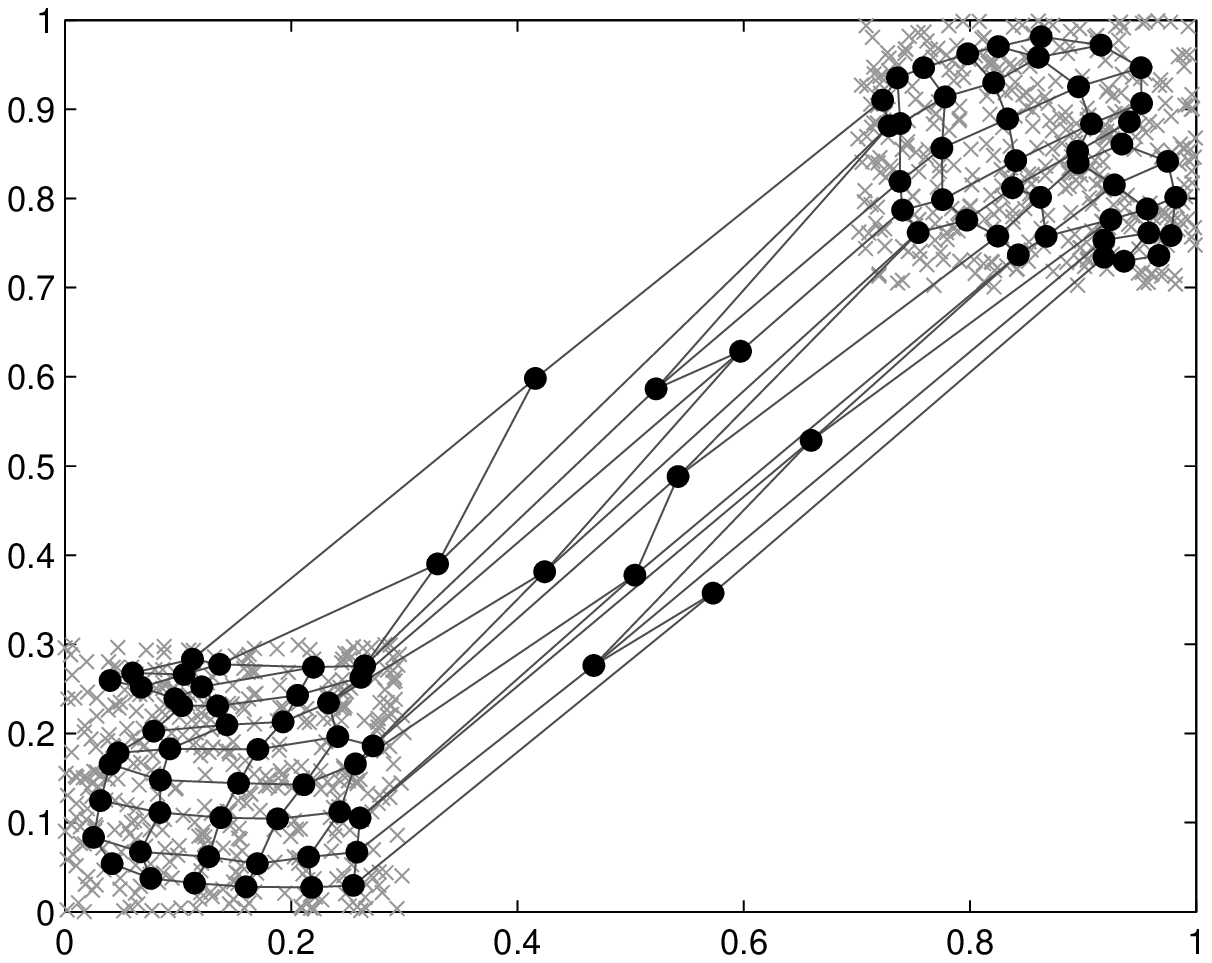} & 
\includegraphics[width=0.50\linewidth]{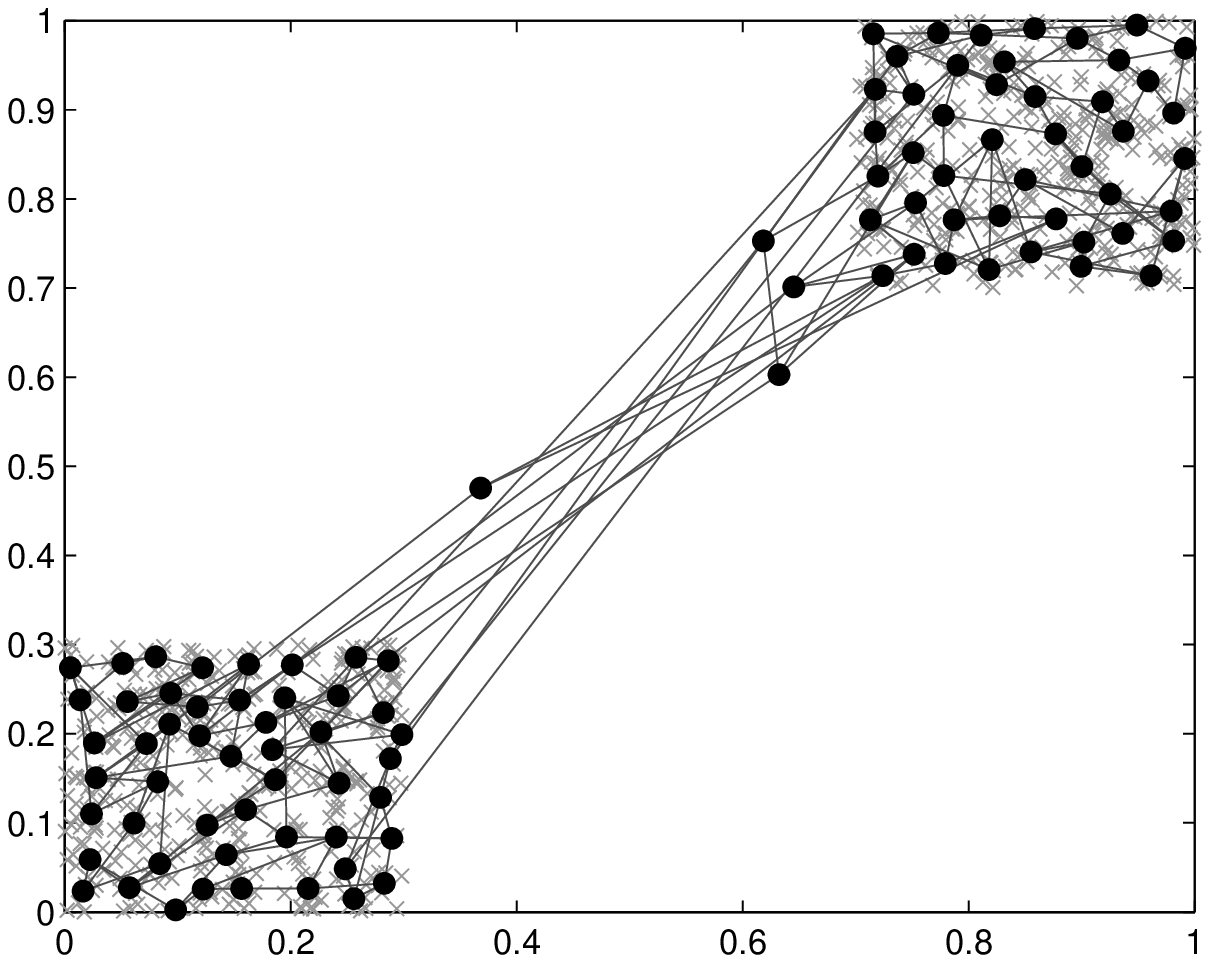}\\
\includegraphics[width=0.50\linewidth]{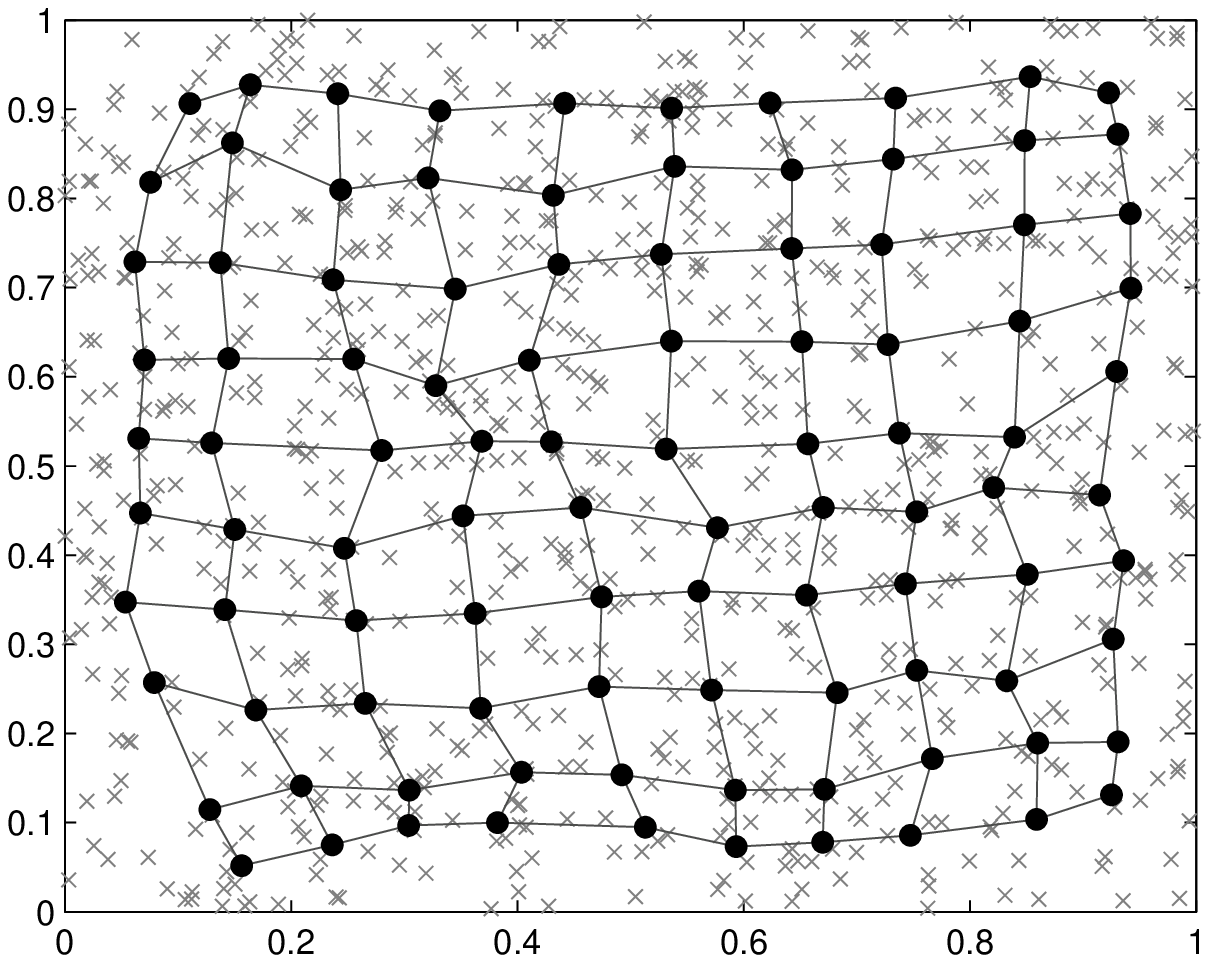} & 
\includegraphics[width=0.50\linewidth]{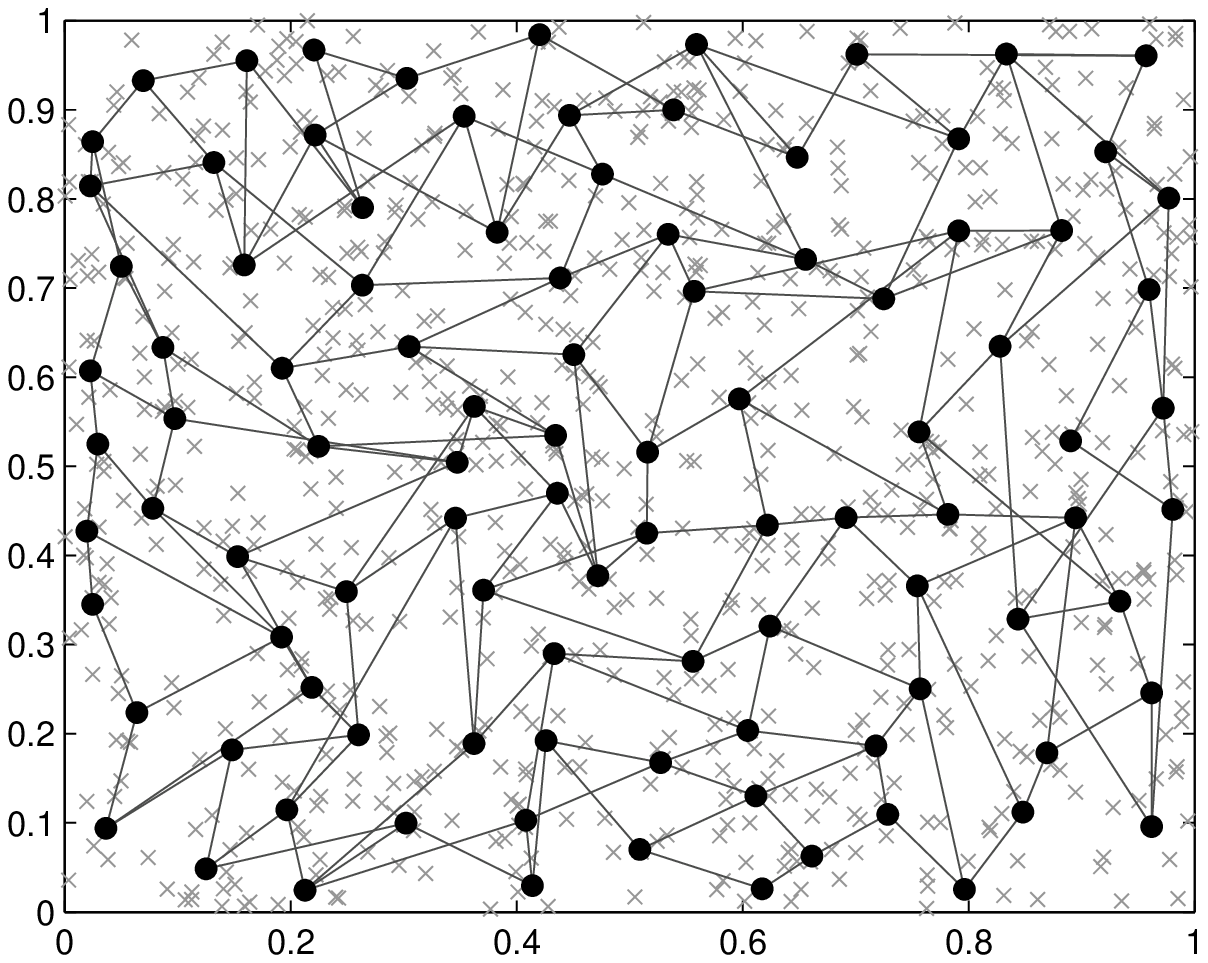}\\
\includegraphics[width=0.50\linewidth]{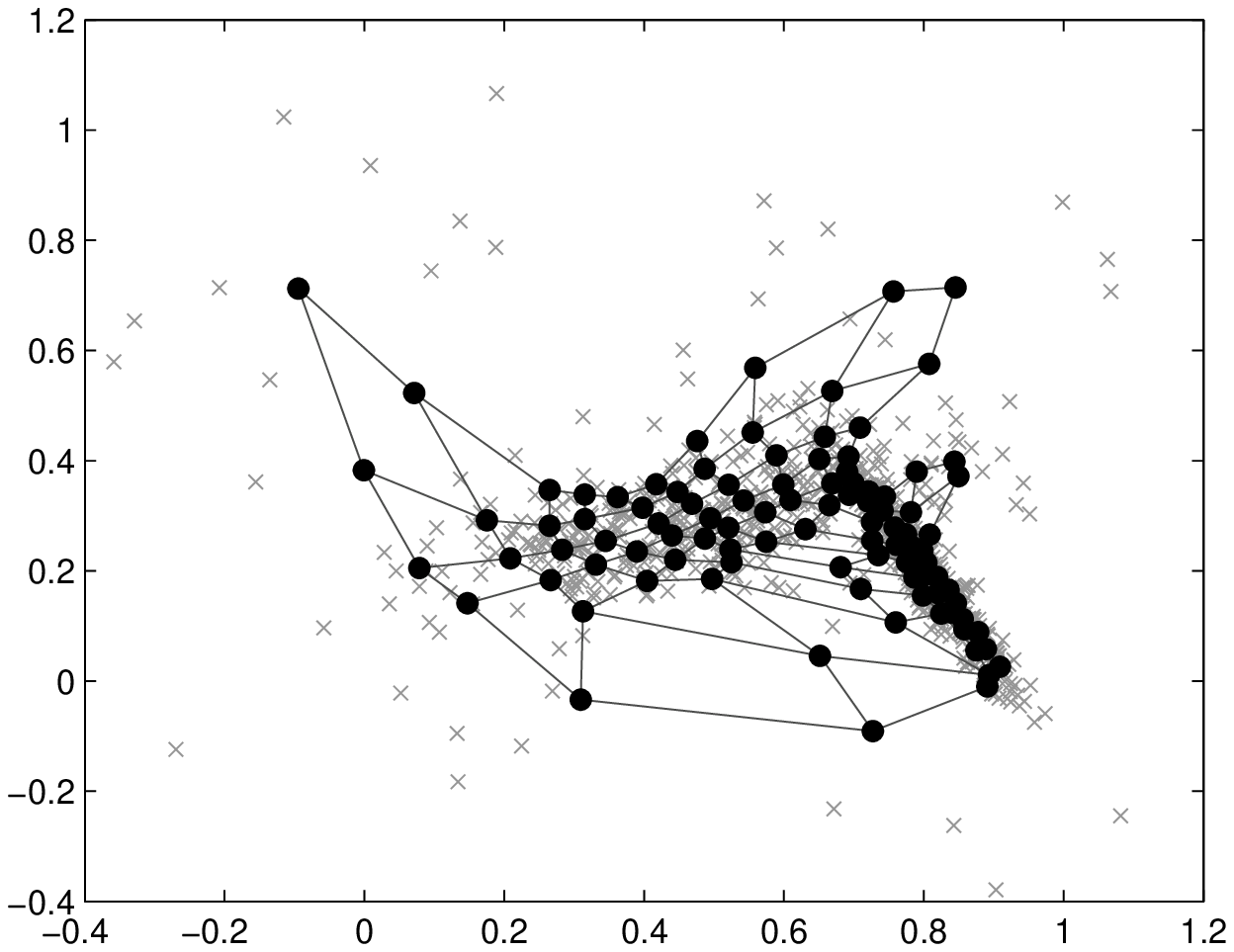} & 
\includegraphics[width=0.50\linewidth]{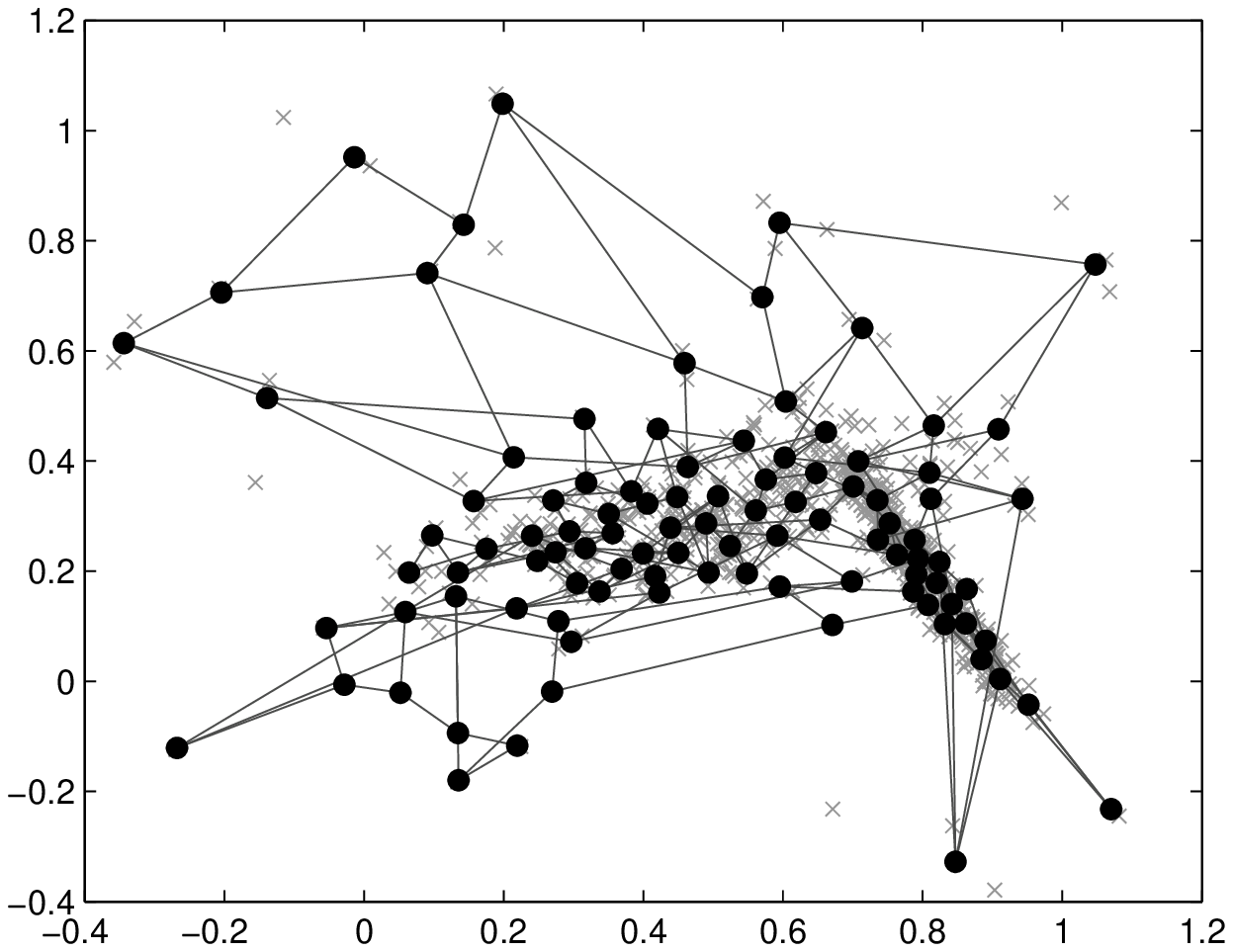}\\
\end{tabular}
\caption{Comparison between standard (left column) and self-instructed multi-rhythm (right column) map, showing stimuli (light crosses) and neurons (bold dots) for 2-D data sets (top=1, middle=2, bottom=3). Lines represent the deformed map at the end of organization.}
\label{fig:comparison}
\end{figure}

{\it Kohonen model}.
The Kohonen network \cite{KohonenStand} is a model of organization of information in the visual cortex in the brain. Information coming from the eye may consist of luminosity/darkness, directions/shapes, colors, motions in all directons of the visual field, represented by a large set of stimuli in a high-dimensional space. One of the brain tasks is to recognize images and group them into classes like, e.g., humans, animals, buildings, roads, etc. In the Kohonen model, this is achieved by reducing (mapping) the high-dimensional manifold of $Q$ stimuli to a low-dimensional (2-D) manifold, represented by $N\ll Q$ neurons, which maintain topological continuity, i.e. proximate stimuli yield proximate neurons. The dynamical rules of the standard Kohonen map are as follows \cite{KohonenStand}. There are stimuli $q=1,\dots,Q$, located at $\vec{p}_{q}$ in a vector space, and there are neurons, $i=1,\dots,N$, living on a 2-D grid map (Fig. \ref{fig:grille10x10}). To each neuron is assigned a weight vector 
$\vec{w}_i$ living in the same vector space as the stimuli. At each time $t$, a stimulus $\vec{p}_{q}$ is randomly selected, and the neuronal weights are updated according to a rule derived from unsupervised Hebbian learning:
\begin{equation}
\vec{w}_{i}(t+1) = \vec{w}_{i}(t) + 
\eta(t)[\vec{p}_{q}(t)-\vec{w}_{i}(t)], ~ 
\forall i\in V_g (t) ~ ,
\label{eq:deltapoids}
\end{equation}
where neuron $g$ is the ``winning'' neuron, being closest to  stimulus $\vec{p}_{q}$, and $V_{g}$ defines an order of neighborhood (topology) for neuron $g$, given by a regular grid map (Fig. \ref{fig:grille10x10}). For example, $V_{g}=1$ corresponds to the 4 nearest neighbors of $g$, while $V_{g}=3$ includes the 20 next-to-nearest neighbors. The parameter $\eta$ denotes the learning rate. Using this rule, the neuronal map learns the topology of the data set and eventually becomes deformed (for example, see Fig. \ref{fig:comparison}).
The order of neighborhood $V(t)$ is initially high, i.e. the map is highly connected. During evolution of organization $V(t)$ decreases gradually, as shown in Fig. \ref{fig:correspondanceVK}. $\eta(t)$ also decreases linearly in time to ensure convergence of neuronal weights.
%

\begin{figure}
\centering
\includegraphics[width=0.9\linewidth]{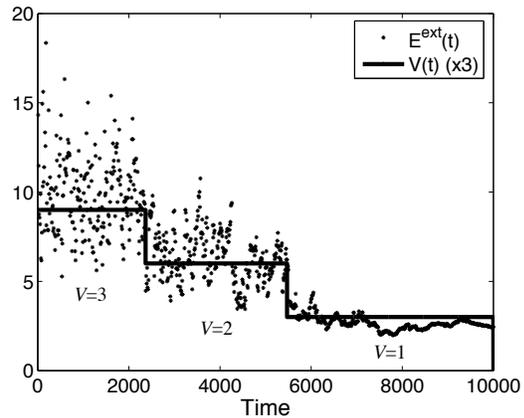}
\caption{Standard Kohonen. Temporal evolution of absolute error $E^{ext}$ and neighborhood order $V$.}
\label{fig:vimpliquee}
\end{figure}
\begin{figure}
\centering
\includegraphics[width=0.9\linewidth]{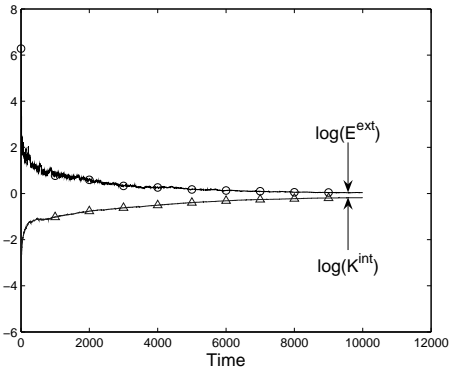}
\caption{Self-instructed multi-rhythm model. 
Correspondence between $K^{int}$ and $E^{ext}$ during the learning process.}
\label{fig:cet}
\end{figure}
\begin{figure}
\centering
\includegraphics[width=0.9\linewidth]{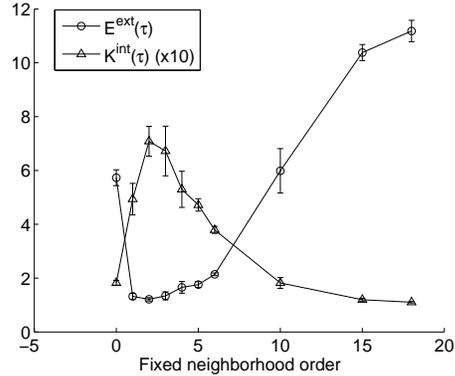}
\caption{Self-instructed multi-rhythm model. 
Correspondence between $K^{int}$ and $E^{ext}$ for given order of neighborhood.}
\label{fig:cegrillefixe}
\end{figure}

{\it Knowledge and auto-adaptivity}.
In the standard Kohonen model, the evolution of the map topology $V(t)$ as well as the learning rate $\eta(t)$ are determined a priori. This conditions the final absolute error of modelization, 
$E^{ext}$, given by 
\be
E^{ext}=\sum_{q=1}^{Q} {\min_{i\in \{1,\ldots,N\}} 
\| \vec{p}_{q}-\vec{w}_{i} \|^{2} }  ~ .
\label{eq:erreurabsolue}
\ee
Fig. \ref{fig:vimpliquee} shows a typical evolution of $E^{ext}$ over time ($V$ goes in steps from 3 to 1).
Guided by the biological principle of adaptation and plasticity of the brain, we suggest to adapt dynamically $V(t)$ and $\eta(t)$ to the state of the network at time $t$, and in particular to the cumulatively learnt information. Moreover, we suggest to introduce a locally independent learning rate $\eta_{i}(t)$ for each neuron. For such purpose we construct (inspired by the Growing Neural Gas network \cite{GNG}) the function of \emph{local attraction} $A^{int}_{i}(t)$, 
\be
A^{int}_{i}(t) = A^{int}_{i}(t-1) + 
\delta_{i,g(t)} \parallel \vec{p}_{q}(t)-\vec{w}_{i}(t)\parallel ~ ,
\label{eq:attractivitelocale}
\ee
for all $i=1,\dots,N$, with initial value $A^{int}_{i}(0)=0$.
This function counts how much the weight vector of a neuron $i$ has moved, being an indicator how much that neuron has learnt.
Low/high attraction means that such neuron has rarely/often been a ``winner''. From local attraction, we construct an adapted learning rate individually different for each neuron (which is biologically more plausible) of the following form
\be
\eta_{i}(t)\propto \frac{1}{A^{int}_{i}(t)} ~ .
\label{eq:tauxdapprentissage}
\ee
The learning rate is chosen to be inversely proportional to local 
attraction, because for a neuron not having learnt at all it is advantageous to have a high potential learning speed, while a neuron which has learnt much tends to saturation, i.e. learning speed becomes slow.
In this way we obtain a new version - called 
\emph{multi-rhythm} - of the Kohonen model. \\
\indent Starting from such new learning rate, we will define an internal 
\emph{knowledge function} $K^{int}$. Intuitively, one expects that learning rates are low and homogeneously distributed among the neurons, when the network has achieved organization, i.e. has learnt much information.
This is the basis for the following definition of the knowledge function 
$K^{int}$, expressed in terms of an harmonic mean,
\begin{equation}
K^{int}(t)=\frac{N}{\displaystyle\sum_{i=1}^N \frac{\overline{A^{int}}(t) }{\overline{A^{int}}(t)  + A^{int}_{i}(t)}} -1  ~ ,
\label{eq:connaissance}
\end{equation}
with initial value $K^{int}(0)=0$ and where $\overline{A^{int}}$ denotes the average of $A^{int}_{i}$ over all neurons $i\in \{1,\ldots,N\}$.
The function $K^{int}$ varies in the range $1/N \le K^{int} \le 1$, and becomes maximal when all $A^{int}_{i}$ are equal.
Fig. \ref{fig:cet} shows the temporal behavior of function $K^{int}$ and the absolute error $E^{ext}$. The decreasing error function $E^{ext}$ mirrors the behavior of function $K^{int}$ and hence justifies the interpretation of the latter as  knowledge gain. Fig. \ref{fig:cegrillefixe} shows the complementary behavior of $E^{ext}$ and $K^{int}$ for the special case where $V(t)$ is kept constant during all of the learning process. It should be noted that although error 
$E^{ext}$ can be used to measure the error at the end of organization, it is not part of the Kohonen network, i.e. not available dynamically. However, 
$K^{int}(t)$ is available to the network at any time $t$. Thus we propose to dynamically adapt the number of connections $M(t)$ using $K^{int}(t)$:
\be
M(t)\propto -K^{int}(t) ~ .
\ee
This yields another version - called \emph{self-instructed} - of the Kohonen model, where the change of topology adapts itself to the rhythm of knowledge gain.
%

\begin{figure}
\centering
\includegraphics[width=0.90\linewidth]{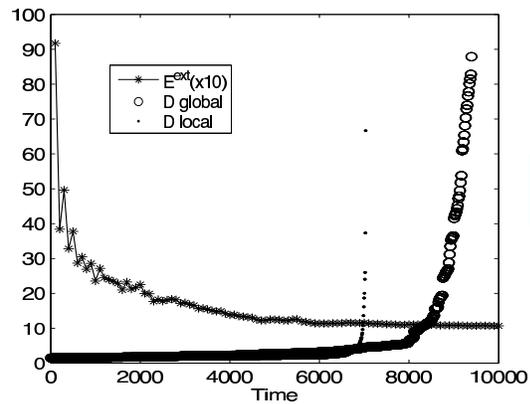}
\caption{Self-instructed multi-rhythm model. 
Global and local connectivity dimension 
$D_{global}$ (open circles), $D_{local}$ (dots) and absolute error 
$E^{ext}$ (asterix) during evolution of organization.}
\label{fig:lienstructurefonction}
\end{figure}
\begin{figure}
\centering
\includegraphics[width=0.95\linewidth]{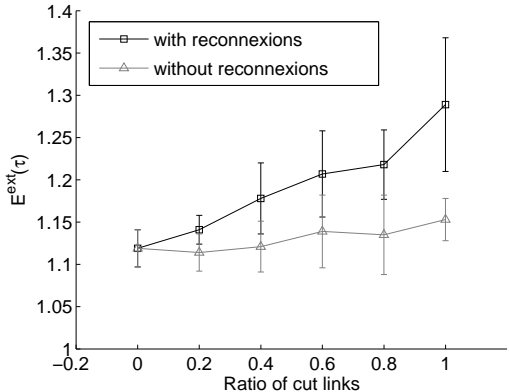}
\caption{Multi-rhythm model. 
Comparison of error for cutting links with and without reconnections. Links cut from the regular grid. Results correspond to data set 2, replicating the test 10 times, but are almost identical for all data sets.}
\label{fig:decrec}
\end{figure}

{\it Dynamical link between topology and information content}.
In information theory, Marchiori and Latora \cite{Marchiori00,Latora01} defined network connectivity dimensions on global and local scales, $D_{global}$ and $D_{local}$, which explicitely show the link between topology and function of the network, which here means efficient transmission of information. A network of SWN architecture is both locally and globally efficient, corresponding to $D_{global}$ and $D_{local}$ being low \cite{Marchiori00,Latora01}.
Fig. \ref{fig:lienstructurefonction} shows the correspondence between $D_{global}$, $D_{local}$, and absolute error $E^{ext}$.
In biology, often there is a link between structure and function, e.g. the structure of the hemoglobin molecule fits its functional task of transporting oxygen. Fig. \ref{fig:lienstructurefonction} shows that $D_{global}$ and $D_{local}$ are both low - indicating SWN connectivity, i.e. the network is highly connected at all length scales. SWN connectivity has been found beneficial in 
supervised learning \cite{Simard05}. 
Here, SWN topology persists during most of the organizational phase, i.e. the regime where the error decreases until saturation. This regime corresponds to an increase of accumulated (learnt) knowledge 
(see Fig. \ref{fig:cet}). Towards the end of organization, the network looses its SWN character (refinement of spatial scale of learning, formation of separate islands of neurons representing classes of stimuli). 
The pruning of connections appears to be an advantage because the progressive independence of neurons leads to a better precision in the local placement of neural weights. This has been confirmed by exploring the alternative of allowing for reconnections, which led to a larger error and deterioration of organization (see Fig. \ref{fig:decrec}). This gives a possible explanation for biologically observed pruning of synaptic connections during organization in the brain of mammals in a relatively short postnatal period. 


{\it Results.}
For a fixed learning duration ($\tau = 10^4$ iterations until stability of organization has been reached), the standard Kohonen model is compared with the self-instructed multi-rhythm version, using a $10\times 10$ neural grid map. First, three different 2-D data sets are tested. They are each composed of $Q=800$ stimuli with different spatial distributions (see Fig. \ref{fig:comparison}). Fig.\ref{fig:comparison} shows that the neurons are closer to stimuli in the self-instructed multi-rhythm version relative to the standard version.  
Also a substantial reduction in the final absolute error $E^{ext}(\tau)$ is observed (see Tab.\ref{tab:standardautoinstruite}). Ranging between only $19\%$ to $66\%$ of the error of the standard Kohonen model, the self-instructed multi-rhythm version is shown to be both more efficient and robust. Indeed, the standard deviations obtained over 10 replications is very low ($\approx1\%$). Second, some high dimensional data sets were also tested. Among them, the ``Pima Indians Diabetes'', 8-dimensions with $Q=768$ stimuli \footnote{\href{http://www.ics.uci.edu/~mlearn/MLRepository.html}{http://www.ics.uci.edu/$\sim $mlearn/MLRepository.html}}, and the ``Hall of Fame'', 15-dimensions with $Q=1320$ \footnote{\href{http://lib.stat.cmu.edu/datasets/}{http://lib.stat.cmu.edu/datasets/}}, and found qualitatively similar results (see Tab. \ref{tab:standardautoinstruite}). 
\begin{table}
\caption{Comparison between standard and self-instructed multi-rhythm model. Absolute error and standard deviation over 10 replications in parentheses.}
\label{tab:standardautoinstruite}
\centering
\begin{tabular}[t]{|c||c|c|c|}
         \hline
         \multirow{2}{*}{\textbf{Data set}} & $\mathbf{E_{ext}(\tau)}$ & $\mathbf{E_{ext}(\tau)}$  & \rule{0pt}{4ex}\multirow{2}{*}{$\mathbf{\frac{E_{ext}(\tau)_{self-instructed}}{E_{ext}(\tau)_{standard}}}$}\\[2ex]
          & \textbf{standard} & \textbf{self-instructed} & \\
		 \hline\hline
         \textbf{1} & $0.446(0.023)$ & $0.222(0.006)$ & $\mathbf{0.498(0.013)}$ \\
         \hline
         \textbf{2} & $1.589(0.029)$ & $1.055(0.014)$ & $\mathbf{0.664(0.009)}$\\
         \hline
         \textbf{3} & $2.302(0.458)$ & $0.442(0.030)$ & $\mathbf{0.192(0.013)}$ \\
		 \hline\hline
         \textbf{Pima} & $5.03(0.37) 10^5$ & $2.33(0.07)10^5$ & $\mathbf{0.462(0.014)}$ \\
         \hline
         \textbf{H. Fame} & $2.06(0.12)10^5$ & $1.156(0.02)10^5$ & $\mathbf{0.562(0.009)}$\\
         \hline
\end{tabular}
\end{table}
%
%

\indent In summary, a dynamical link between geometry and function (organization) was introduced in the Kohonen network, based on its dynamically aquired information. Thus adaptation, also found in nature, leads to substantial improvement of organization in the context of our model. At the beginning of organization, the SWN topology causes communication at all scales, both global and local, and the network learns very quickly. We postulate that such topology is beneficial in any information organizing network. As knowledge increases, the communication scale changes towards more local connectivity, until total independence of neurons due to pruning is reached. We found pruning of connections (in contrast to reconnections) to play a beneficial role in this model of self-organization 
of the visual cortex. \\
%
\indent This work was supported by NSERC Canada grants to H.K. and M.P.


\end{document}